\documentclass{Pos}

\usepackage{graphicx}
\usepackage{subfig}
\usepackage{amsmath}

\title{Measurement of the $\pmb{\Lambda}$(1405) in proton proton reactions with HADES}

\ShortTitle{Reconstruction of the $\Lambda(1405)$}

\author{\speaker{Johannes Siebenson}\thanks{Laura Fabbietti, Alexander Schmah, Eliane Epple}\\
        HADES Collaboration\\
        E-mail: \email{johannes.siebenson@ph.tum.de}}

\abstract{We present an analysis of the $\Lambda(1405)$ resonance in $p+p$ reactions at a kinetic beam energy of $3.5$ GeV, measured by the \textbf{H}igh \textbf{A}cceptance \textbf{D}i-\textbf{E}lectron \textbf{S}pectrometer (HADES). The resonance is reconstructed in the two charged decay channels $\Sigma^{\pm}\pi^{\mp}$, with help of a kinematic refit, which improves the mass resolution. The high misidentification of pions and protons as kaons required the development of a sophisticated sideband analysis, which can describe the misidentification background quite well. 
}

\FullConference{XLVIII International Winter Meeting on Nuclear Physics, BORMIO2010\\
		January 25-29, 2010\\
		Bormio, Italy}

\begin{document}

\section{Motivation}
Although the $\Lambda(1405)$ is known since several decades, its inner structure is still a topic of ongoing investigation. For instance, in \cite{Weise,Oset} the $\Lambda(1405)$ is described as a dynamically generated resonance appearing in the scattering theory with coupled meson-baryon channels of strangeness $S = -1$. In this approach, the $\Lambda(1405)$ is actually a superposition of two poles in the scattering amplitude, one with a strong coupling to a $\Sigma\pi$ resonance and one dominated by a quasi-bound state of a $K^-p$ system. As the two poles appear quite close in energy, only a single resonance can be observed experimentally. With the prediction of deeply bound kaonic states \cite{Yamazaki}, a different, more phenomenological approach was supposed, in which the $\Lambda(1405)$ has a single-pole character, generated only by a bound state of a kaon and a nucleon. This triggered further theoretical and experimental efforts. Indeed, assuming that the $K^-p$ state dominates the resonance, the production of a bound state like $ppK^-$ could proceed through the $\Lambda(1405)$ doorway. However, the lack of high statistics and good quality data for the $\Lambda(1405)$ made a verification of these different theoretical approaches (single- or double-pole character) difficult in the last years. Additionally, a full understanding of the resonance is only possible, if it is reconstructed in different experiments, since theory predicts the $\Lambda(1405)$ line shape to be sensitive to the initial reaction. Measurements with kaon and pion beams are hampered by low statistics \cite{Hemingway,Thomas}. Recent progress was made in the sector of $\gamma$ induced reactions \cite{CLAS}, and data from $p+p$ reactions were published, which show the decay $\Lambda(1405)\rightarrow\Sigma^0+\pi^0$ \cite{ANKE}. However a complete picture of the $\Lambda(1405)$ requires also the investigation of the other decay branches. \\
We have investigated $p+p$ reactions at $3.5$ GeV kinetic beam energy with HADES at GSI and attempted to reconstruct the $\Lambda(1405)$ by its decay into the channels $\Sigma^{\pm}\pi^{\mp}$, which is the first measurement in this reaction system and this decay branch.

\section{The HADES experiment}
The experiment was performed with the \textbf{H}igh \textbf{A}ccaptance \textbf{D}i- \textbf{E}lectron \textbf{S}pectrometer (HADES) at the heavy-ion synchrotron SIS at GSI Helmholtzzentrum für Schwerionenforschung in Darmstadt, Germany. A detailed description of the spectrometer can be found in \cite{HADES}. HADES consists of a $6$-coil toroidal magnet, centered on the beam axis, and six identical detection sections, located between the coils and covering polar angles from $18^\circ$ and $85^\circ$. In the measurement, presented here, the six sectors are comprised of a gaseous Ring-Imaging Cherenkov (RICH) detector, four planes of Multi-wire Drift Chambers (MDCs) for track reconstruction, and two Time-Of-Flight walls (TOF and TOFino), supplemented at forward polar angles with Pre-Shower chambers. In forward direction a Forward Wall (FW) hodoscope, consisting of scintillating plastic cells, was implemented for the first time in HADES. It covers a polar angular range from $0.33^\circ$ to $7^\circ$. However, the data of the FW were so far not used for the analysis presented here.\\
A proton beam of $3.5$ GeV kinetic energy was incident on a $LH_2$ target. During the beam time $1.2\cdot10^9$ triggered events were collected.
\\
Hadrons were identified by the combined measurement of their momentum, reconstructed with the MDCs ($\frac{\Delta p}{p} \approx 1-5\%$) and their energy loss ($dE/dx$) in the MDCs or in the time-of-flight detectors. An additional independent identification is possible by the measurement of the velocity ($\beta$) of the particles together with the momentum ($p$). The mass of the hadrons can then be determined according to $m_0=p/(\beta\gamma c)$.

\section{The analysis of the $\pmb{\Lambda(1405)}$}
\begin{figure}[h]
	\centering
		\includegraphics[width=0.60\textwidth]{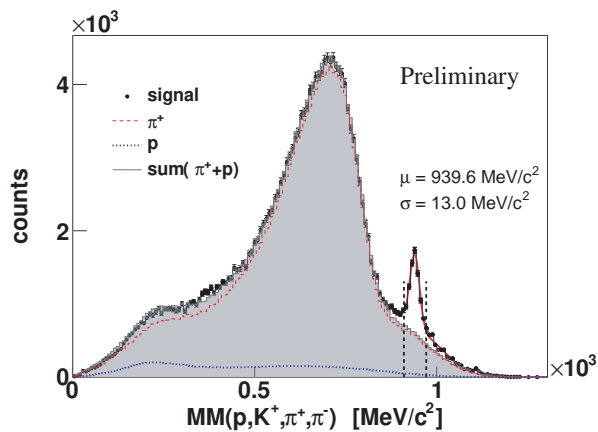}
	\caption{Missing mass distribution ($p,K^+,\pi^+,\pi^-$) showing a missing neutron. The background (gray distribution) is obtained with a sideband analysis on the kaon mass.}
	\label{fig:Neutron}
\end{figure}

The $\Lambda(1405)$ is produced in $p+p$ reactions together with a $K^{+}$ and a proton. Thus it can be reconstructed via the missing mass technique. A characteristic feature of the $\Lambda(1405)$ analysis is that it has to take into account the effect of a nearby resonance, the $\Sigma(1385)^0$, which has almost the same event characteristic. As both resonances are broad, the $\Sigma(1385)^0$ will overlap the $\Lambda(1405)$ in the missing mass ($p,K^+$) spectrum. Thus, in order to get a pure $\Lambda(1405)$ signal, the $\Sigma(1385)^0$ together with other contributions has to be subtracted from the spectra.
\\The investigated decay chains of the $\Lambda(1405)$ read as follows:
\begin{eqnarray}
\Lambda(1405)\rightarrow\Sigma^{\pm}+\pi^{\mp}\rightarrow (n+\pi^{\pm})+\pi^{\mp} \label{Chain}
\end{eqnarray}
Both channels provide two additional charged particles ($\pi^+,\pi^-$), which can be identified and allow to do an exclusive analysis. 
After the identification of the appropriate set of hadrons ($p,K^+,\pi^+,\pi^-$), the first step in the analysis is to investigate the missing mass of these particles, as shown in figure \ref{fig:Neutron}. As expected from reaction (\ref{Chain}), a neutron peak is clearly visible. The remainder of the spectrum is only due to the misidentification of protons and pions as $K^+$. This misidentification background can be modeled with a sophisticated sideband analysis on the kaon mass \cite{Siebenson}. For that purpose two additional data samples are produced, one containing only protons, explicitly misidentified as kaons, and the other one containing only pions, explicitly misidentified as kaons. This can be done by shifting the mass cut, previously used to identify the true $K^+$, either to higher or lower masses. These sideband samples are analyzed in the same way as the true data. The result is shown in the gray histogram in figure \ref{fig:Neutron}, which is the sum of the contributions due to misidentified pions (red dashed) and protons (blue dotted).
 
\begin{figure}[h]
   \centering
   \subfloat[]{\includegraphics[width=0.49\textwidth]{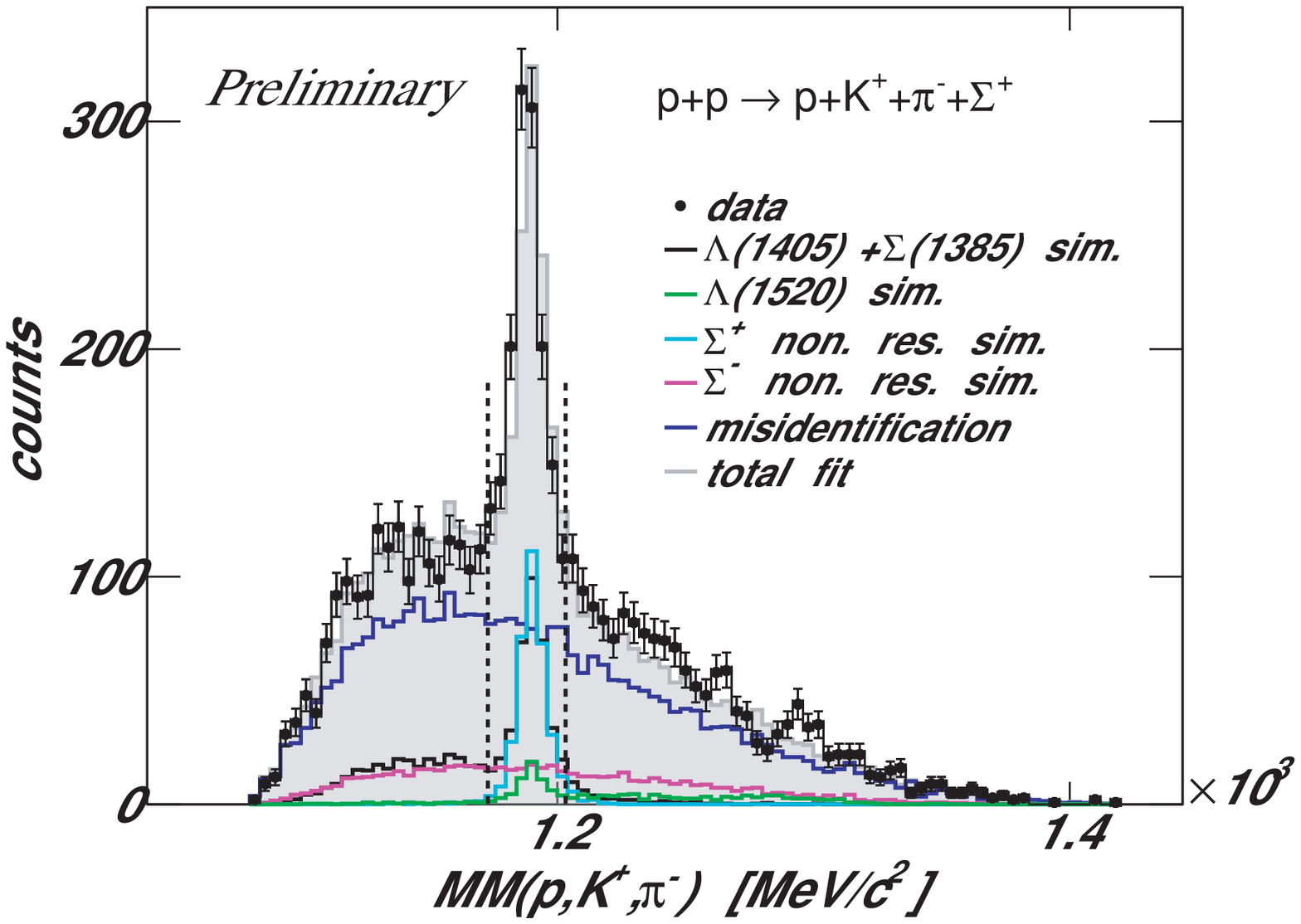}}
      \hspace{0.0\textwidth}
   \subfloat[]{\includegraphics[width=0.49\textwidth]{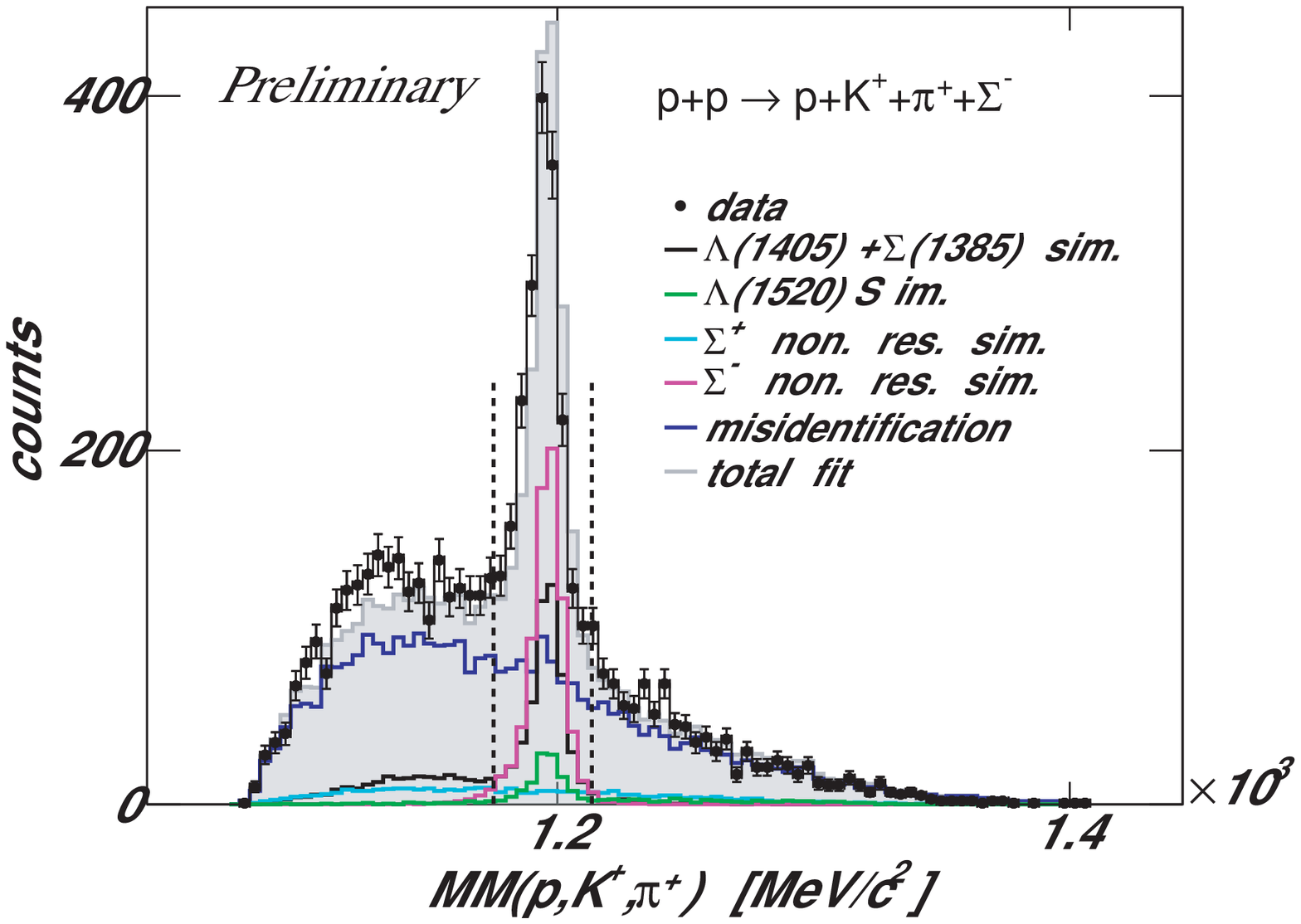}}
   \caption{(a): Missing mass of ($p,K^+,\pi^-$) showing the $\Sigma^+$ signal. (b): Missing mass of ($p,K^+,\pi^+$) showing the $\Sigma^-$ signal. Events within the vertical dashed lines ($3\sigma$) are identified as being part of decay channel ($\Sigma^+\pi^-$) (a) or ($\Sigma^-\pi^+$) (b), respectively. }
\label{fig:SigmaPM}
\end{figure}

\begin{figure}[h]
   \centering
   \subfloat[]{\includegraphics[width=0.49\textwidth]{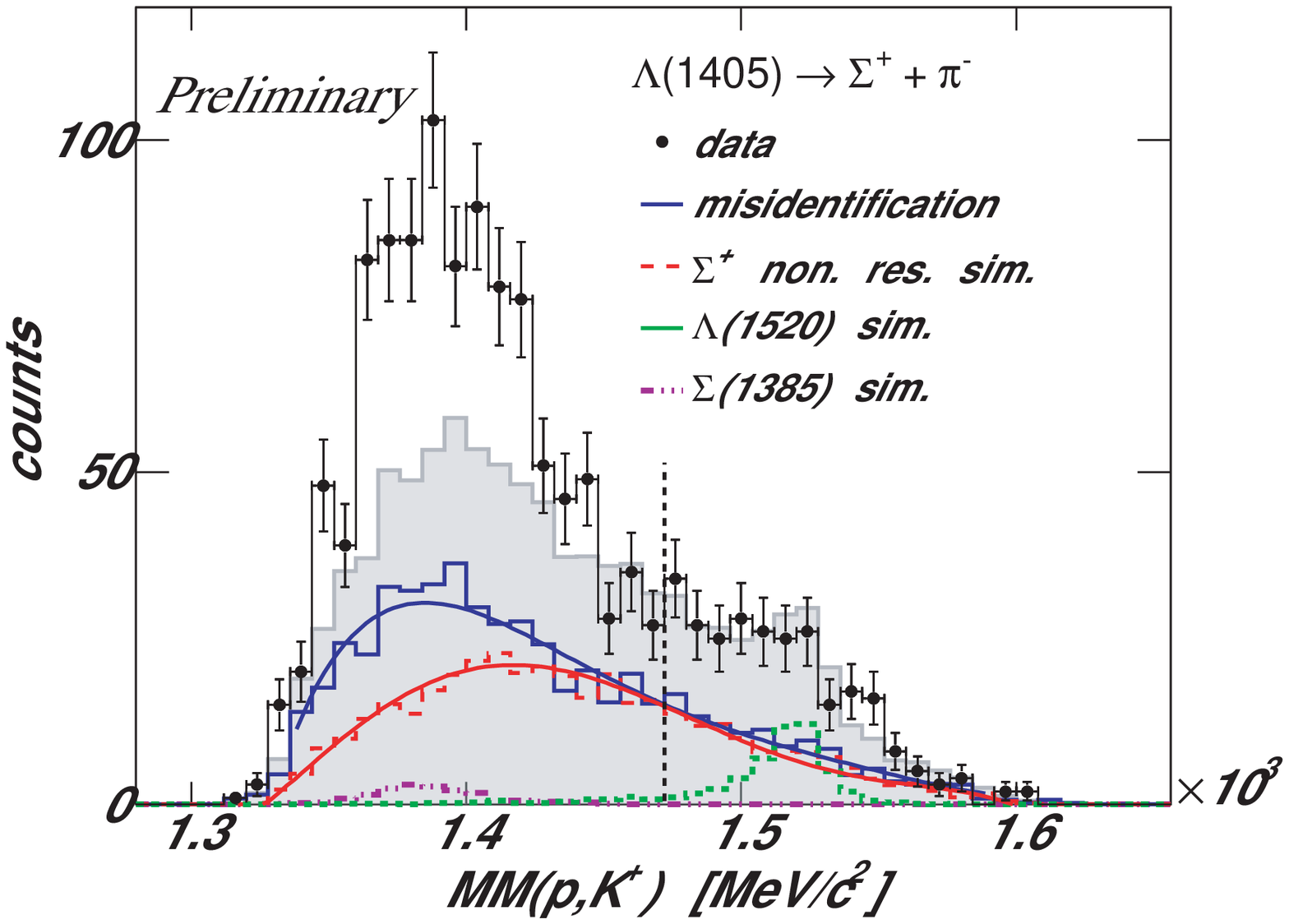}}
      \hspace{0.0\textwidth}
   \subfloat[]{\includegraphics[width=0.49\textwidth]{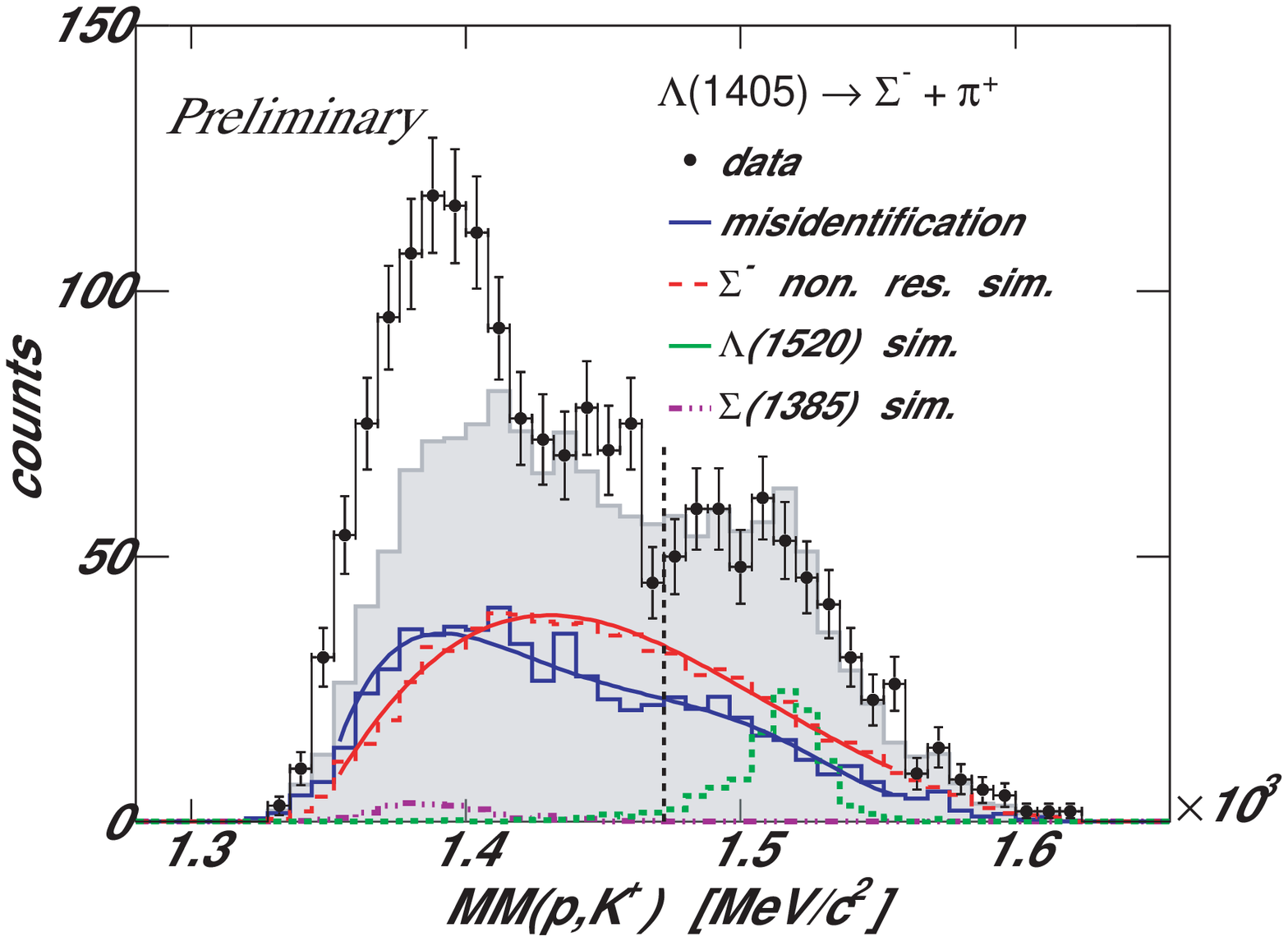}}
   \caption{(a): Missing mass of ($p,K^+$) for the $\Sigma^+\pi^-$ decay channel. (b): Missing mass of ($p,K^+$) for the $\Sigma^-\pi^+$ decay channel. The fits to the non-resonant distributions (red histograms), the fits to the misidentification background (blue histograms) and the $\Sigma(1385)^0$ (violet histograms) can be subtracted in order to obtain the pure $\Lambda(1405)$.}
\label{fig:La1405PM}
\end{figure}

\noindent The obtained distribution can describe the complete spectrum quite well. The relative contributions of pions and protons to the total background is determined via $\chi^2$ minimization, where the ratio between the red dashed and the blue dotted histogram is the fit parameter and the summed spectra (gray histogram) is adapted to the black data points within the range from $0$ to $860$ MeV/c$^2$. \\
A further event selection is done by cutting on a mass range from $909$ MeV/c$^2$ to $970$ MeV/c$^2$ ($2.4\sigma$), between the vertical dashed lines (see figure \ref{fig:Neutron}). These events are kinematically refitted, with the constraint that the missing mass of all particles must be the neutron mass. The kinematic refit is a well established tool in the investigation of elementary particle reactions, since it can improve the mass resolution of particles, reconstructed via invariant- or missing mass technique \cite{Avery}. The refitted tracks are used to calculate the missing mass of ($p,K^+,\pi^-$) and ($p,K^+,\pi^+$), exhibited in figure \ref{fig:SigmaPM} (black data points). In both spectra, a clear signal due to $\Sigma^+$ (panel (a)) and $\Sigma^-$ (panel (b)) is visible, as it is expected from reaction (\ref{Chain}).
By cutting on the appropriate mass range between the vertical dashed lines ($3\sigma$ cut), events are selected which should include the contribution from the $\Lambda(1405)$ resonance. Furthermore, the data sample can be separated in this way into the two different decay channels ($\Sigma^+\pi^-$) or ($\Sigma^-\pi^+$). The final spectra of the missing mass ($p,K^+$) are plotted in figure \ref{fig:La1405PM} (black points), where picture (a) refers to the decay $\Lambda(1405)\rightarrow\Sigma^++\pi^-$ and picture (b) refers to $\Lambda(1405)\rightarrow\Sigma^-+\pi^+$. 


\noindent Figure \ref{fig:SigmaPM} as well as figure \ref{fig:La1405PM} show additionally the different contributions to the experimental spectra. Besides the $\Lambda(1405)$ and $\Sigma(1385)^0$, there are a couple of other channels, which contaminate the signal. The non-resonant reactions $pp\rightarrow pK^{+}\Sigma^+\pi^-$ or $pp\rightarrow pK^{+}\Sigma^-\pi^+$ are described by simulations. They deliver the main contribution to the peaks in figure \ref{fig:SigmaPM} (light blue and magenta histograms) and create a broad phase space distribution in figure \ref{fig:La1405PM} (red dashed histograms). 
The $\Lambda(1520)$ (dark green), which is also obtained with simulations, has the same event characteristic as the $\Lambda(1405)$ but the overlap of both resonances in the missing mass ($p,K^+$) spectra is low. A large background source in all pictures is due to the misidentification (dark blue histograms), obtained by analyzing the sideband sample, mentioned above.\\ 
All these contributions are fitted simultaneously to the full mass range of figure \ref{fig:SigmaPM} and to the right tails of figure \ref{fig:La1405PM} (above the vertical dashed lines), since there is no $\Lambda(1405)$ signal expected. \\
As seen in figure \ref{fig:La1405PM} the sum of the background contributions reproduces quite well the region on the right side of the $\Lambda(1405)$ signal. The differences in the background distributions in figure \ref{fig:La1405PM} a) and b) are on the one hand due to the cuts, applied so far, and on the other hand due to the different cross sections of the non-resonant $\Sigma^-\pi^+$ and $\Sigma^+\pi^-$ production. Since the background shape has been deduced from simulations, further systematic studies and calculations with higher statistics must be carried out before the pure $\Lambda(1405)$ signal and its spectral distribution become accessible. The present data sample is promising in this respect.  

\end{document}